\documentstyle[aps,preprint,graphicx,floats]{revtex}

\def\gs{g_S^{}}

\begin{document} 

\marginparwidth 1.cm
\setlength{\hoffset}{-1cm}
\newcommand{\mpar}[1]{{\marginpar{\hbadness10000%
                      \sloppy\hfuzz10pt\boldmath\bf\footnotesize#1}}%
                      \typeout{marginpar: #1}\ignorespaces}
\def\mda{\mpar{\hfil$\downarrow$\hfil}\ignorespaces}
\def\mua{\mpar{\hfil$\uparrow$\hfil}\ignorespaces}
\def\mla{\marginpar[\boldmath\hfil$\rightarrow$\hfil]%
                   {\boldmath\hfil$\leftarrow $\hfil}%
                    \typeout{marginpar: $\leftrightarrow$}\ignorespaces}

\def\ba{\begin{eqnarray}}
\def\ea{\end{eqnarray}}
\def\bq{\begin{equation}}
\def\eq{\end{equation}}

\renewcommand{\abstractname}{Abstract}
\renewcommand{\figurename}{Figure}
\renewcommand{\refname}{Bibliography}

\newcommand{\eg}{{\it e.g.}\;}
\newcommand{\ie}{{\it i.e.}\;}
\newcommand{\etal}{{\it et al.}\;}
\newcommand{\ibid}{{\it ibid.}\;}

\newcommand{\mx}{M_{\rm SUSY}}
\newcommand{\pt}{p_{\rm T}}
\newcommand{\et}{E_{\rm T}}
\newcommand{\del}{\varepsilon}
\newcommand{\sla}[1]{/\!\!\!#1}
\newcommand{\fb}{{\rm fb}}
\newcommand{\gev}{{\rm GeV}}
\newcommand{\tev}{{\rm TeV}}
\newcommand{\abi}{\;{\rm ab}^{-1}}
\newcommand{\fbi}{\;{\rm fb}^{-1}}

\newcommand{\zpc}[3]{${\rm Z. Phys.}$ {\bf C#1} (#2) #3}
\newcommand{\epc}[3]{${\rm Eur. Phys. J.}$ {\bf C#1} (#2) #3}
\newcommand{\npb}[3]{${\rm Nucl. Phys.}$ {\bf B#1} (#2)~#3}
\newcommand{\plb}[3]{${\rm Phys. Lett.}$ {\bf B#1} (#2) #3}
\renewcommand{\prd}[3]{${\rm Phys. Rev.}$ {\bf D#1} (#2) #3}
\renewcommand{\prl}[3]{${\rm Phys. Rev. Lett.}$ {\bf #1} (#2) #3}
\newcommand{\prep}[3]{${\rm Phys. Rep.}$ {\bf #1} (#2) #3}
\newcommand{\fp}[3]{${\rm Fortschr. Phys.}$ {\bf #1} (#2) #3}
\newcommand{\nc}[3]{${\rm Nuovo Cimento}$ {\bf #1} (#2) #3}
\newcommand{\ijmp}[3]{${\rm Int. J. Mod. Phys.}$ {\bf #1} (#2) #3}
\renewcommand{\jcp}[3]{${\rm J. Comp. Phys.}$ {\bf #1} (#2) #3}
\newcommand{\ptp}[3]{${\rm Prog. Theo. Phys.}$ {\bf #1} (#2) #3}
\newcommand{\sjnp}[3]{${\rm Sov. J. Nucl. Phys.}$ {\bf #1} (#2) #3}
\newcommand{\cpc}[3]{${\rm Comp. Phys. Commun.}$ {\bf #1} (#2) #3}
\newcommand{\mpl}[3]{${\rm Mod. Phys. Lett.}$ {\bf #1} (#2) #3}
\newcommand{\cmp}[3]{${\rm Commun. Math. Phys.}$ {\bf #1} (#2) #3}
\newcommand{\jmp}[3]{${\rm J. Math. Phys.}$ {\bf #1} (#2) #3}
\newcommand{\nim}[3]{${\rm Nucl. Instr. Meth.}$ {\bf #1} (#2) #3}
\newcommand{\prev}[3]{${\rm Phys. Rev.}$ {\bf #1} (#2) #3}
\newcommand{\el}[3]{${\rm Europhysics Letters}$ {\bf #1} (#2) #3}
\renewcommand{\ap}[3]{${\rm Ann. of~Phys.}$ {\bf #1} (#2) #3}
\newcommand{\jhep}[3]{${\rm JHEP}$ {\bf #1} (#2) #3}
\newcommand{\jetp}[3]{${\rm JETP}$ {\bf #1} (#2) #3}
\newcommand{\jetpl}[3]{${\rm JETP Lett.}$ {\bf #1} (#2) #3}
\newcommand{\acpp}[3]{${\rm Acta Physica Polonica}$ {\bf #1} (#2) #3}
\newcommand{\science}[3]{${\rm Science}$ {\bf #1} (#2) #3}
\newcommand{\vj}[4]{${\rm #1~}$ {\bf #2} (#3) #4}
\newcommand{\ej}[3]{${\bf #1}$ (#2) #3}
\newcommand{\vjs}[2]{${\rm #1~}$ {\bf #2}}
\newcommand{\hep}[1]{${\tt hep\!-\!ph/}$ {#1}}
\newcommand{\hex}[1]{${\tt hep\!-\!ex/}$ {#1}}
\newcommand{\desy}[1]{${\rm DESY-}${#1}}
\newcommand{\cern}[2]{${\rm CERN-TH}${#1}/{#2}}

\preprint{
 {\vbox{
 \hbox{MADPH--02--1318}
 \hbox{DESY--03--002}
 \hbox{hep-ph/0301039}}}
}

 
\title{TeV resonances in top physics at the LHC}

\author{T. Han$^a$, D. Rainwater$^b$ and G. Valencia$^c$}

\address{$^a$Department of Physics, University of Wisconsin, Madison, WI 53706\\
$^b$Theory Group, DESY, Hamburg, Germany\\
$^c$Department of Physics and Astronomy, Iowa State University,
Ames, IA 50011}

\maketitle

\begin{abstract} 
  
  We consider the possibility of studying novel particles at the TeV
  scale with enhanced couplings to the top quark via top quark pair
  production at the LHC and VLHC. In particular we discuss the case of
  neutral scalar and vector resonances associated with a strongly
  interacting electroweak symmetry breaking sector. We constrain the
  couplings of these resonances by imposing appropriate partial wave
  unitarity conditions and known low energy constraints. We evaluate
  the new physics signals via $WW\to t\bar t$ for various models
  without making approximation for the initial state $W$ bosons, and
  optimize the acceptance cuts for the signal observation. We conclude
  that QCD backgrounds overwhelm the signals in both the LHC and a
  200~TeV VLHC, making it impossible to study this type of physics in
  the $t\bar{t}$ channel at those machines.

\end{abstract} 


\tightenlines

\section{Introduction}  

Elucidating the mechanism of electroweak symmetry breaking constitutes
a top priority for the next generation of collider experiments. A
possibility that remains open is that in which there are no light
Higgs bosons and the interactions among the longitudinal vector bosons
become strong at a scale of ${\cal O}$(1 TeV) where new dynamics must
set in~\cite{sews}. This possibility has regained interest in light of
certain inconsistencies associated with the forward-backward asymmetry
$A^b_{FB}$ measured at LEP~\cite{chano}. Whereas the leptonic
quantities prefer a low Higgs mass, the hadronic quantities prefer a
high Higgs mass with $M_H$ as large as 1.25~TeV allowed at the 90\%
confidence level \cite{chano,chano2}. Even if one disregards the
$A^b_{FB}$ problem, there are known scenarios in which a heavy Higgs
can be consistent with precision electroweak data~\cite{peswells}.
Finally there is also the possibility of having a heavy composite
Higgs boson~\cite{chivukula} which has not been excluded by the
precision data.

The phenomenology of strongly interacting electroweak symmetry
breaking has been studied at length.  In particular, it has been
established that it is possible to extract signals for a heavy Higgs
boson or for a techni-rho-like resonance by studying the processes
$WW\to WW$ and $WZ \to WZ$ at the LHC~\cite{baggeretal}.  One sector
that has not been studied completely consists of new resonances with
couplings to fermions that prefer the third generation.  This is an
interesting possibility given that the top quark mass is very close to
the electroweak scale. Much theoretical work has been carried out
recently in connection with the top quark and the electroweak
sector~\cite{topcolor,topss,liz}. Some early phenomenological
attempts have been made to evaluate the signal rates at future
hadron colliders \cite{cpetal} and at linear colliders \cite{lcs,us}.
In a previous paper we have studied in detail 
the phenomenological implications of new TeV resonances and their
couplings to the top quark in connection with future lepton
colliders~\cite{us}.  With the relatively clean experimental environment,
a high energy $e^+e^-$ linear collider will have significant sensitivity to probe the 
possible new TeV resonances that couple strongly to the top-quark sector.
What is missing is to quantitatively analyze the observability of this class
of signals at the LHC.   

In this paper we extend the study to the LHC
and consider the possibility of observing a signal in the $WW \to
t\bar{t}$ channel over the large QCD background.
Of particular importance to this study will be the application of
recently established techniques to suppress the QCD backgrounds that
plague this process~\cite{WBF2}. Our intent is to study these
backgrounds in sufficient detail to establish whether the LHC has any
chance of studying these signal processes. In fact, a quantitative conclusion
in this regard has been long overdue.

\section{Model for the Resonances}\label{sec:model}

We wish to establish the level at which the LHC can probe couplings of
heavy new resonances to the top quark. We have in mind possible
resonances associated with a strongly interacting electroweak symmetry
breaking sector in which the top quark plays a special role (TOP-SEWS).
However, for this study we will not consider any specific model of
this type but rather we will discuss generic resonances with masses of
order 1~TeV.

To write down the couplings of these resonances to standard model (SM)
particles we employ the framework of effective Lagrangians. We start
with the SM in the limit of an infinitely heavy Higgs boson. This
scenario has been studied extensively in the literature~\cite{heavyh,bessre}
so we will not repeat it here. As is well known~\cite{unit}, this
non-renormalizable model gives rise to amplitudes for longitudinal $W$
and $Z$ scattering that violate unitarity at energies near a TeV. It
is expected that new physics appearing at this scale modifies these
amplitudes and restores unitarity.

In the scenario we have in mind, the lightest degrees of freedom
associated with the new physics are the resonances of interest. In
particular we consider two cases: a heavy neutral scalar (similar to a
heavy Higgs); and a triplet of heavy vectors (similar to a color-singlet
techni-rho triplet). Within the framework of an effective theory, these degrees
of freedom will be responsible for restoration of unitarity in the
longitudinal $W$ and $Z$ scattering amplitudes at first.  Of course,
our model including these resonances, remains an effective
non-renormalizable theory and unitarity will still be violated. We
will use this feature to select our otherwise arbitrary couplings. In
the same spirit of studies of strong $W$, $Z$ scattering at the
SSC~\cite{oldssc}, we will choose our couplings so that all amplitudes
satisfy simple unitarity constraints up to the highest possible
energy. We will find that this energy is of order 3~TeV, and is
sufficiently high to remain outside the range probed by the LHC. In
this manner, our study remains independent of dynamical details
responsible for restoration of unitarity.

The effective Lagrangians describing all these interactions are the
same ones discussed in Ref.~\cite{us}. However, here it will be
important to calculate the SM electroweak and QCD backgrounds and the
$WW\to t\bar t$ signal exactly (without
recourse to the Equivalence Theorem or the effective $W$
approximation). To calculate the signals simultaneously, we will need
the couplings of the new degrees of freedom to the physical SM gauge
bosons and we present those here.

First we need the interactions between the SM gauge bosons and a
generic scalar resonance $S$, as have been considered in
Ref.~\cite{baggeretal}. The leading order effective Lagrangian for
these interactions contains two free parameters: the resonance mass
$M_S$ and a coupling constant $\gs$ that can be traded for the width
of the new resonance into $W$ and $Z$ pairs. The resulting coupling
(for $V^\mu = W^{\pm \mu},Z^\mu$) is
\begin{equation}
{\cal L}(SV^\mu V^\nu) = \gs g M_W g^{\mu\nu}. 
\label{elsr}
\end{equation}
This coupling is proportional to the $HV^\mu V^\nu$ coupling in the SM,   
and produces a width into $W$ and $Z$ pairs in the limit $M_S \gg M_W$ 
given by,
\begin{equation}
\Gamma_{Sww} = {3\over 32\pi}{g_S^2 M_S^3 \over v^2}.
\label{swlwl}
\end{equation}
With $\gs=1$, Eq.~(\ref{swlwl}) reproduces the width of 
the SM Higgs boson into longitudinal $W$ and $Z$ pairs.

Similarly, the effective interaction between the top quark and the
scalar resonance is described by the Lagrangian,
\begin{equation}
{\cal L}= - \kappa {m_t\over v} S\bar t t.
\label{lstt}
\end{equation}
The new coupling $\kappa$ can be traded for the width of the scalar
into top quark pairs,
\begin{equation}
\Gamma_{St\bar{t}} = {3\kappa^2\over 8\pi}{m_t^2 M_S \over v^2}
\left(1-{4m_t^2\over M_S^2}\right)^{3\over 2}\; .
\label{stt}
\end{equation}
Once again, the case $\kappa=1$ corresponds to the SM Higgs boson
coupling to the top quark.

The interactions of SM gauge bosons with new vector resonances have
also been described in the literature~\cite{bagsm,bess} (dubbed as
the BESS model, Breaking the Electroweak Symmetry Strongly~\cite{bess}). 
In the notation of Ref.~\cite{bagsm}, two new parameters are introduced, $a$
and $\tilde{g}$ (these correspond to $\alpha$ and $g''/2$ from
Ref.~\cite{bess}, respectively). Working in the limit in which the
resonance is much heavier than the $W$ and $Z$, we only need the
effective interaction between one heavy vector $V$ and two SM gauge
bosons. For the neutral $V$ it is given by (with all momenta
incoming),
\begin{equation}
{\cal L}(V^{0\lambda}(p_3)W^{-\mu}(p_1)W^{+\nu}(p_2) =
-{g^2\over 4\tilde{g}}\bigl[(p_1-p_2)_\lambda g_{\mu\nu}+ 
(p_2-p_3)_\mu g_{\nu\lambda} + (p_3-p_1)_\nu g_{\mu\lambda}\bigr]\; .
\label{vecel}
\end{equation}

The use of this vertex facilitates the numerical implementation of the
signal obtained using the Equivalence Theorem
(Goldstone-bosons $w^{\pm},z$ equivalent to the longitudinal
gauge bosons at high energies \cite{unit}). However, we are not
computing all the effects of electroweak strength that would appear in
a full implementation of the BESS model \cite{bessre}. 
The constant $\tilde{g}$ is related to the mass and width of the
vector particles $V$ via the relation,
\begin{equation}
\Gamma(V^0 \to w^+ w^-) = {M_V^5 \over 192 \pi v^4 \tilde{g}^2}\; .
\end{equation}

Finally we consider effective couplings between the new vector mesons
and the top and bottom quarks. We limit our study to direct
non-universal couplings because we are interested in potentially large
effects from the top quark in electroweak symmetry breaking. This
implies that we want to explore couplings significantly larger than
those induced by mixing between $V^0$ and the neutral SM gauge bosons.
Severe low energy constraints on such couplings for light fermions
(the $b$ parameter in the BESS model \cite{bessre}), also force us to
consider only non-universal couplings specific to the top (and maybe
bottom) quark. In view of this we write the effective interaction
\begin{equation}
{\cal L}_{eff} = 
 - \bar \psi \gamma^\mu (g_V+g_A\gamma_5)\tau^i \/\psi\ V^i_\mu\;  ,
\label{ttrho}
\end{equation}
for $\bar\psi$ being the third generation quark doublet
$(\bar{t}\;\bar{b})$. These effective couplings result in the partial
width
\begin{equation}
\Gamma(V^0 \rightarrow  t \bar{t}) = {M_V \over 4 \pi}
\left(1-4{m_t^2\over M_V^2}\right)^{1/2} \left[ g_V^2
\left(1+2{m_t^2\over M_V^2} \right)+g_A^2\left( 1 + 
4{m_t^2\over M_V^2} \right)\right]\; .
\end{equation}
In accordance with the discussion for lepton colliders in Ref.~\cite{us}, 
we consider the case 
\begin{equation}
g_A = -g_V\;, \;\;\;~|g_V| \leq 0.03 |\tilde{g}|\,
\label{veclec}
\end{equation} 
in order to satisfy the low energy constraints. 
The first condition ensures that there are no large right-handed 
couplings in the  $Wtb$ vertex, which are 
severely constrained by $b\rightarrow s \gamma$. 
The second condition follows from the one-loop contributions 
of new left-handed $Wtb$ couplings to the $Z \rightarrow b\bar{b}$ 
partial width. 

\section{Unitarity constraints}

To choose the parameters $\gs$, $\kappa$ and $\tilde{g}$ of
Eqs.~(\ref{elsr}), (\ref{lstt}) and (\ref{vecel}), we adopt the strategy of
using these resonances to increase as much as possible
the energy at which perturbative
unitarity is violated by tree-level amplitudes in our model.
The aim of this approach is to generate simple models of
resonances which do not result in unphysically large predictions for
event rates at the LHC.  We consider unitarity conditions for those
channels which yield the strongest constraints as established in the
literature \cite{unit,mvw}.

For the scalar resonance we use two conditions. First, we require that
the magnitude of the $J=0$, $I=0$ partial wave for $w^\pm,z$
scattering remain below its unitarity bound, $|a^0_0| \leq
1$ \cite{unit}, at least up to about $\sqrt{s} = 4$~TeV. We are not
concerned with violations of this condition at higher $WW$ center of
mass energy because the LHC (and VLHC) cannot produce observable event
rates in that region. This condition will constrain the allowed values
of $\gs$ for a given value of $M_S$.  Second, we require that the
process $ww\rightarrow t\bar{t}$ satisfy inelastic partial-wave
unitarity in the $I=0,J=0$, color singlet channel as in
Ref.~\cite{mvw}, $a_0(ww\rightarrow t\bar{t})_{I=0,{\rm singlet}} <
1/2$.  Once again, we require this condition to be valid for $WW$
center of mass energy up to about 4~TeV, and this constrains the
possible values of $\kappa$. In both cases we apply the unitarity
conditions to the scattering of pseudo-Goldstone bosons, invoking the
Equivalence Theorem~\cite{unit}.  We require that the two conditions
be satisfied below the resonance, in the resonance region, and as far
above the resonance region as possible.  To satisfy perturbative
unitarity in the resonance region, it is necessary to treat the
resonance width as in Ref.~\cite{seymour}.

\begin{figure}[t]
\begin{center}
\includegraphics[width=14cm]{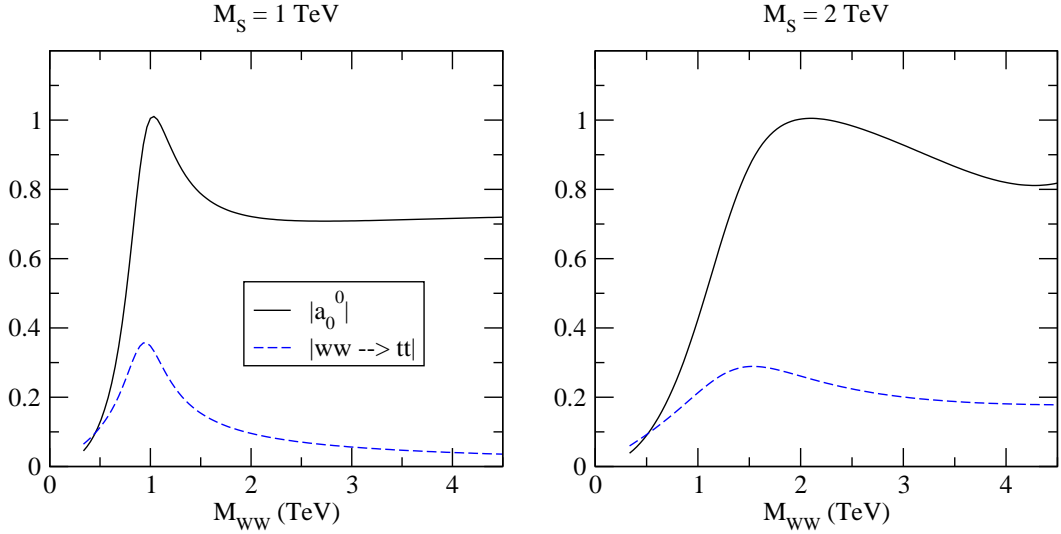}
\end{center}
\caption{Unitarity constraints for heavy Scalars. In (a) we show a Higgs-like 
  1~TeV scalar with $\Gamma(S \rightarrow WW) = 493$~GeV and $\Gamma(S
  \rightarrow t \bar{t}) = 50$~GeV, whereas in (b) we show a model
  with a broad enhancement in the $I=J=0$ channel corresponding to
  parameters $M_S = 2$~TeV, $\gs = 0.9$ and $\kappa = 1.35$. In both
  cases perturbative unitarity is satisfied up to 4~TeV.}
\label{fig:scalar}
\end{figure}

Using the amplitudes obtained in Ref.~\cite{us} and the prescription
in Ref.~\cite{seymour}, the two conditions become:
\begin{eqnarray}
|a^0_0(s)| &=& \left|-{\Gamma_{Sww}\over g_S^2 M_S}
{s\left(1+i{\Gamma \over M_S}\right) \over s-M_S^2+ i\Gamma_{Sww} {s\over M_S} }
\biggl(1+{s\over M_S^2}(g_S^2-1)\biggr) \right. \nonumber \\ 
&+& \left. {\Gamma_{Sww}\over 3 g_S^2 M_S} \biggl(
{s\over M_S^2}(g_S^2-1)-2g_S^2\biggl[1-{M_S^2\over s}\log
\biggl(1+{s\over M_S^2}\biggr)\biggr]\biggr)\right| \; < 1
\label{a0scalar}
\end{eqnarray}
for the $I=J=0$ partial wave amplitude for $WW$ scattering, and 
\begin{equation}
 {3G_Fm_t\sqrt{s}\over 8\pi\sqrt{2}}\ \biggl|
{M_S^2+s(\gs\kappa -1)\over s-M_S^2}\biggr|<0.5\; .
\label{wwttunit}
\end{equation}
Examples of parameters that satisfy these unitarity constraints for
$M_S=1,\; 2$~TeV are shown in Fig.~\ref{fig:scalar}. For a 1~TeV
scalar we have chosen parameters $\gs=\kappa=1$ which reproduce the SM
Higgs widths $\Gamma(S \rightarrow ww) = 493$~GeV and $\Gamma(S
\rightarrow t \bar{t}) = 50$~GeV. For the 2~TeV scalar we have chosen
$\gs = 0.9$ and $\kappa = 1.35$ which correspond to $\Gamma(S
\rightarrow ww) = 3.2$~TeV and $\Gamma(S \rightarrow t \bar{t}) =
210$~GeV respectively. Clearly, an object with such a large ``width''
cannot be called a resonance, and we simply interpret it as an example
of a non-resonant model with a broad enhancement in the $I=J=0$ $WW$
scattering channel in the spirit of older such models
\cite{unit,oldssc}. We note that it is not possible to choose a narrow
scalar resonance with mass $M_S = 2$~TeV because the $I=J=0$ partial
wave violates the unitarity condition at very low energy.  It is easy
to understand this if we recall that we are using this scalar
resonance to ``fix'' the violation of unitarity that occurs in this
partial wave for the universal low energy theorems \cite{oldssc} at
$1.2$~TeV. Clearly if the resonance occurs far from this energy and it
is too narrow, it will not have sufficient impact to delay the
violation of unitarity. This tells us that it is possible to have a
heavy ($\sim 2$~TeV) and narrow scalar resonance only if there is
additional physics that is responsible for restoration of partial wave
unitarity.

\begin{figure}[t]
\begin{center}
\includegraphics[width=14cm]{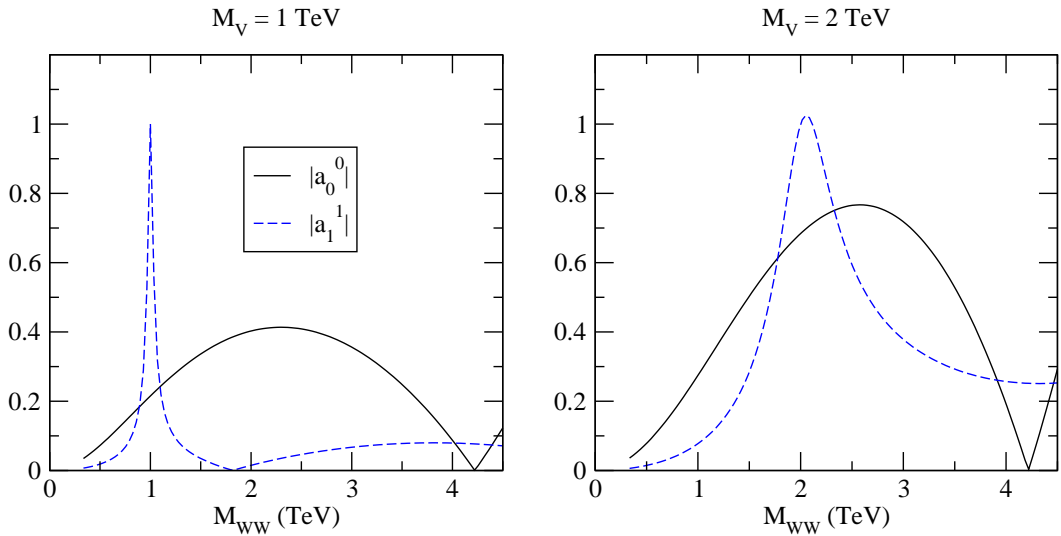}
\end{center}
\caption{Unitarity constraints for heavy Vectors. In (a) we show a 
  1~TeV vector with $\Gamma(V \rightarrow WW) = 50$~GeV and $\Gamma(V
  \rightarrow t \bar{t}) = 1.3$~GeV, whereas in (b) we show a 2~TeV
  vector with $\Gamma(V \rightarrow WW) = 500$~GeV and $\Gamma(V
  \rightarrow t\bar{t}) = 8.2$~GeV. In both cases perturbative
  unitarity is satisfied up to 4~TeV.}
\label{fig:vector}
\end{figure}

For the vector resonance the inelastic process $ww\to t\bar{t}$
satisfies the unitarity constraint in all cases once we impose the low
energy constraint, Eq.~(\ref{veclec}). However, there are two partial
waves in $ww$ scattering that could be problematic~\cite{vecuni} in
this case. We demand that $|a^0_0|\leq 1$ and that $|a^1_1|\leq 1$,
which using the Equivalence Theorem correspond to:
\begin{equation}
|a^0_0(s)| = \left|{s\over 16 \pi v^2}+6{\Gamma_W \over M_V}
\left[\left(2+{M_V^2\over s}\right)\log\left(1+{s\over M_V^2}\right)
-1-{3s\over 2M_V^2}\right]\right|\; < \;1\;,
\label{a0vector}
\end{equation}
and 
\begin{eqnarray}
|a^1_1(s)| &=& \left| {s\over 96 \pi v^2} + i\,{\Gamma_W^2 s \over (s-M_V^2)+
\Gamma_W^2 M_V^2} 
+ 3{\Gamma_W \over M_V}\left(1+2{M_V^2\over s}\right)
\left(2+{M_V^2\over s}\right)\log\left(1+{s\over M_V^2}\right) \right.
\nonumber \\
&-&
\left. {{\Gamma_W \over M_V}\over (s-M_V^2)+\Gamma_W^2 M_V^2} 
\left[6{M_V^6 \over s} +{3s^3\over 2 M_V^2}+10s^2-{35\over 2}sM_V^2
\right]+\cdots \right|\; < \;1.
\label{a1vector}
\end{eqnarray}
In this second expression we have not written down terms of higher
order in $\Gamma_W/M_V$, which we have denoted by the ellipses.
Examples of parameters satisfying both conditions for a 1~and~2~TeV
vector resonance are shown in Fig.~\ref{fig:vector}.  For the 1~TeV
resonance we use $\tilde{g} = 3$ and $g_V = -g_A = 0.09$ which
correspond to $\Gamma(V \rightarrow WW) = 50$~GeV and $\Gamma(V
\rightarrow t\bar{t}) = 1.3$~GeV. For the case $M_V = 2$~TeV we choose
$\tilde{g} = 5.38$ and $g_V = -g_A = 0.16$, corresponding to $\Gamma(V
\rightarrow WW) = 500$~GeV and $\Gamma(V \rightarrow t\bar{t}) =
8.2$~GeV respectively. We have chosen both of these resonances to be
relatively narrow. Notice also that the width into $t\bar{t}$ induced
by the direct coupling is almost two orders of magnitude larger than
the one induced by mixing. Finally we point out that the vector is
sufficiently narrow to simply use a constant width for the $s$-channel
vector propagator.

\begin{figure}[t]
\begin{center}
\includegraphics[width=14cm]{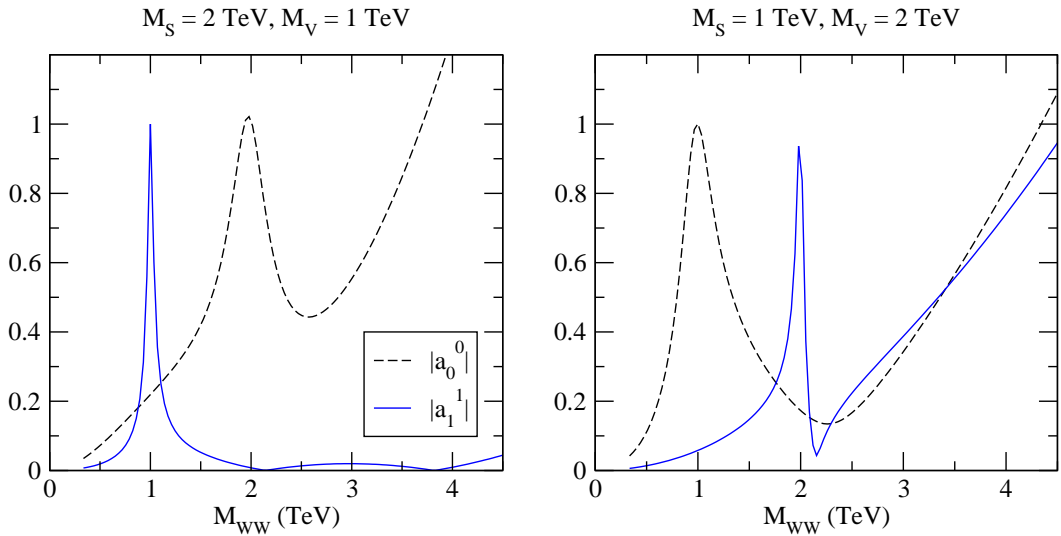}
\end{center}
\caption{Models with two resonances that satisfy the perturbative unitarity 
  constraints with parameters described in the text.}
\label{fig:scavec}
\end{figure}

As argued above, it is not possible to have a heavy and narrow scalar
unless there are additional ingredients in the model. Simple models
that can be constructed may include additional light scalars or
additional resonances. Here we present two simple examples including
only one scalar and one vector resonance.  In Fig.~\ref{fig:scavec} we
illustrate two choices of parameters that satisfy all the unitarity
conditions that we have discussed. The first one has a 2~TeV scalar
with a 1~TeV vector and couplings $\gs = 0.3$, $\tilde{g} = 3$,
$\kappa = 1$ and $g_V = 0.09$. This translates into $\Gamma(S
\rightarrow WW) = 355$~GeV, $\Gamma(V \rightarrow WW) = 50$~GeV,
$\Gamma(S \rightarrow t\bar{t}) = 114$~GeV, and $\Gamma(V \rightarrow
t\bar{t}) = 1.3$~GeV.  In the second example we chose $M_S=1$~TeV and
$M_V=2$~TeV with $\gs = 0.8$, $\tilde{g} = 15.5$, $\kappa = 1$ and
$g_V = 0.43$. This leads to $\Gamma(S \rightarrow WW) = 315$~GeV,
$\Gamma(V \rightarrow WW) = 60$~GeV, $\Gamma(S \rightarrow t\bar{t}) =
50$~GeV, and $\Gamma(V \rightarrow t\bar{t}) = 60$~GeV.

\section{Sensitivity to the Top-SEWS Signal at the LHC}

Clearly, one direct way of probing enhanced electroweak couplings to
the top quark at the LHC is $t\bar{t}$ production via weak boson
fusion. Although gluon fusion via heavy quark loops typically leads to
a larger production rate for a scalar resonance, the gauge boson
fusion processes may be more advantageous because of the enhanced
gauge boson coupling to new physics, as well as advantages of the
final state kinematics, which have been used previously in studies of
intermediate mass and heavy Higgs bosons at hadron
colliders~\cite{WBF2,wpwp,baggeretal}. In fact, there is no viable 
technique to make use of the gluon-fusion
channel for a heavy Higgs signal $H \to t\bar t$ due to the overwhelming 
background $gg, q\bar q \to t\bar t$. We thus concentrate on the
$W$-fusion channel and explore its observability. 
For a careful study of the background we must
include all EW Feynman graphs for pair production of two top quarks
and two additional light partons, $qq'\to q''q'''t\bar{t}$.

We first consider SM heavy Higgs production and decay to a top quark
pair, for Higgs boson mass values $M_H = 360,800$~GeV. This provides a
baseline for comparison of non-SM scenarios with SM expectations.
Next we consider two different heavy non-SM scalar resonances,
$M_S = 1.0,2.0$~TeV, for which the Feynman diagrams mentioned above
apply, but with enhanced couplings (magnitude) of the scalar to weak
bosons and top quarks as discussed in Sec.~\ref{sec:model}. We then
consider production of vector states for $M_V = 1.0,2.0$~TeV, for
which the Feynman diagrams still contain a Higgs scalar with
$M_H\to\infty$ (set to $10^8$~GeV in practice) in addition to the new
vector resonance graphs, and finally the 2~TeV vector case of the last
model considered, consisting of both a heavy non-SM scalar and vector.
The $M_H\to\infty$ terms are used to reproduce the low energy theorems
of the effective theory and are only needed in the case of the heavy
vector. Our formulation of the heavy scalar already includes those
terms (also applied to the last model).

The major background to any of these potential signals is SM QCD
production of $t\bar{t}$ plus two partons, ``$t\bar{t}jj$'', which has
been described elsewhere~\cite{WBF2} as one of the dominant SM
backgrounds that LHC Higgs searches will face. 
Before stronger cuts are imposed on the
forward partons, a total cross section for this background is not easy
to obtain, as the calculation experiences soft and collinear
singularities of the additional partons. The limiting case would of
course be the total $t\bar{t}$ cross section, where the emission of
additional soft or collinear partons is not observed.

Table~\ref{tab:rates} shows the model parameters used for each
resonance considered, their resulting total widths, LHC production
cross sections without kinematic cuts or decays, and cross sections
with decays and cuts required for detector observation (discussed in
detail below). Notice that the widths in this Table do not agree
exactly with what we say in Sec.~\ref{sec:model}. The differences
arise from the additional electroweak strength terms in the numerical
implementation as compared to the widths calculated in
Sec.~\ref{sec:model} with the aid of the Equivalence Theorem. 
Our numerical estimates do not use the equivalence theorem. For the 
scalar resonance they correspond to an exact tree-level calculation with
the model defined by Eqs.~\ref{elsr}~and~\ref{lstt}. 
For the vector resonance we assume 
that the direct $s$-channel exchange of the new resonance dominates, and that 
the direct coupling of the resonance to $t \bar{t}$ is larger than the 
mixing-induced coupling.
For the cross section without decays or cuts, we restrict the
kinematic range of the two forward scattered quarks to $p_T > 10$~GeV,
to avoid collinear singularities in the background calculations.

\begin{table}
\caption{Model parameters, total widths and resultant total cross
sections for the SM Higgs boson, non-SM scalar and non-SM vector
resonances considered. Parameters are chosen to maximize the cross
section without violating the unitarity constraints derived in
Sec.~\ref{sec:model}. In addition we show the QCD $t\bar{t}jj$
continuum. All cross sections are without decay of the top quarks.
We include kinematic cuts on the two scattered far forward scattered
partons as described in the text, to make the calculations stable
against soft and collinear singularities. The QCD top quark background is uncertain
to a factor of several at this stage, due to the scattered partons
being well into the non-perturbative region of QCD for this level of
cuts.}  
\label{tab:rates}
\vskip 3.mm
\begin{tabular}{c|c|c|c|c|c|c}
state & $M$ (GeV) & $\gs,\kappa$ & $\tilde{g},g_V$ &
$\Gamma$ (GeV) & $\sigma_{``tot''}$ (fb) & $\sigma_{dec/cuts}$ (fb) \\
\tableline
SM $H$ &  360 & -,-       & -,-       &   17.6 & 29.0 & 1.40 \\
SM $H$ &  800 & -,-       & -,-       &  308   & 25.3 & 1.56 \\
$S$    & 1000 & 1.0 ,1.0  & -,-       &  519   & 21.2 & 1.25 \\
$S$    & 2000 & 0.9 ,1.35 & -,-       & 3339   & 16.5 & 0.86 \\
$V$    & 1000 & -,-       & 3.0 ,0.09 &   51.6 & 18.0 & 0.97 \\
$V$    & 2000 & -,-       & 5.38,0.16 &  507   & 13.2 & 0.84 \\
$S+V$  & 1000,2000 & 0.8,1.0 & 15.5,0.43 & 349,119 & 20.5 & 1.33 \\
\tableline
QCD $t\bar{t}jj$ & - & -,- & -,-  & - & ~few $\times 10^5$ & 7200
\end{tabular}
\end{table}

Unfortunately, the background cross section is several orders of magnitude 
larger than any of the signal processes we consider. 
To determine if there is any hope of extracting a signal from this
extremely large QCD continuum, we must consider specific final state
signatures. Dual hadronic decay of the top quarks constitutes the
largest branching ratio, about $45\%$, and yields an eight jet final
state, of which two (typically central) jets are bottom quarks, and
two light jets are required to be found very far forward.
Unfortunately there is a question as to whether such all-jet final
states can be accepted by the trigger due to their large rate, so we
do not consider this channel. Instead, we analyze semi-leptonic decay
of the top quark pair, which has a $29\%$ branching ratio and yields
two far forward/backward light jets, two central $b$ jets, two central
light jets which could be reconstructed to a $W$ boson, a central
lepton (only $e$ or $\mu$ is considered), and significant missing
transverse energy. To make the simulation more realistic we impose
Gaussian smearing to the final state particles' momenta according to
CMS expectations~\cite{CMS}.

For this final state we impose first the minimum set of kinematic cuts
that will be required for detector acceptance and identification of
the final state:
\begin{eqnarray}
\label{eq:cuts1}
\nonumber
& p_T(j) > 20~{\rm GeV} , \qquad p_T(b) > 20~{\rm GeV} , \qquad 
p_T(\ell) > 15~{\rm GeV} , \\
& |\eta(j)| < 4.5 \; , \qquad \qquad |\eta(b)| < 2.5 \; , \qquad \qquad
|\eta(\ell)| < 2.5 \; ,    \\
\nonumber
&
\Delta R(jj,jb,bb) > 0.4 \; , \qquad 
\Delta R(j \ell) > 0.4 \; , \qquad 
\Delta R(\ell \ell) > 0.2 \; , \\
\nonumber
& \sla{p}_T > 30~{\rm GeV} .
\end{eqnarray}
We also immediately apply the basic ``rapidity gap'' cuts well
established in~\cite{WBF2}, which reduces the various signals by a
typical factor 4, and the background in this case by about an order of
magnitude:
\begin{eqnarray}
\label{eq:cuts2}
& \eta_{j,min} + 0.6 < \eta_{\ell_{1,2}} < \eta_{j,max} - 0.6 \, ,
\nonumber \\
& \eta_{j_1} \cdot \eta_{j_2} < 0 \, , \qquad 
\triangle \eta_{tags} = |\eta_{j_1}-\eta_{j_2}| \geq 4.2 \, ,
\end{eqnarray}
leaving a gap of at least 3 units of pseudo-rapidity in which the
charged lepton, $b$ jets and $W$ boson decay jets can be observed.
Given four jets in the final state, not counting additional QCD
radiation, there must be some criteria to select which two will be the
forward tagging jets. We find that to a very high efficiency,
typically on the order of $90\%$ or better, selection of the two most
energetic jets fulfills this requirement. If instead the two highest
$p_T$ jets are chosen, the cross section drops by a factor of four,
due to one of the jets from the hadronic $W$ boson being chosen as a
tagging jet candidate. The event in those cases rarely satisfies the
requirement that all other final state particles lie in the rapidity
region between the tagging jets. With these cuts imposed, the various
signals have resulting cross sections all at around the 1~fb level, as
shown in Table~\ref{tab:rates}. However, the QCD background is a
factor 7000 larger. Clearly this is far too much to have any hope of
observing any of the signals.

An estimate of the reduction factor necessary for signal observation
can be obtained by calculating the maximum number of background events
allowable for, say, a $5\sigma$ statistical significance, given the
likely number of signal events in 300~fb$^{-1}$ of data. To be
realistic we apply two factors of 0.5 for the probability to tag the
two $b$ quark jets, two factors of $68\%$ for the expected efficiency
to identify the forward tagging jets, and a $90\%$ lepton ID
efficiency. We ignore the loss of signal that comes from
reconstructing the hadronic $W$ boson, or its parent top quark (one
might argue this is not necessary given the complex final state
signature). This leaves us with on the order of 30 signal events, or
an allowable background of about 40 events -- a factor 6000 reduction
of the QCD background.

\begin{figure}[!htb]
\begin{center}
\includegraphics[width=10cm,angle=90]{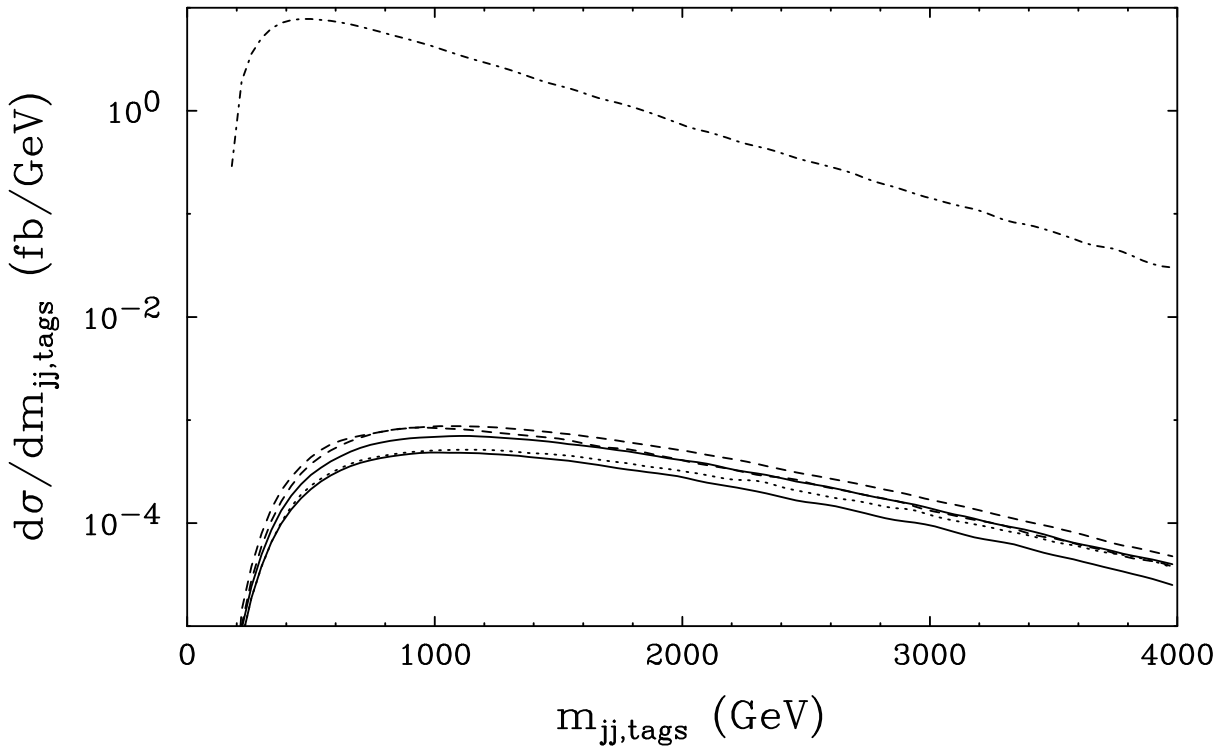}
\end{center}
\caption{Invariant mass of the forward tagging jets at the LHC, selected
  as the two highest energy jets in the event. The SM Higgs boson
  signals for $M_H = 360,800$~GeV are shown by the dashed curves, the
  two non-SM scalar cases by the solid curves, the non-SM heavy
  vectors by the dashed curves, and the QCD continuum by the dot-dashed
  curve.}
\label{fig:mtags_l}
\end{figure}
\begin{figure}[!htb]
\begin{center}
  \includegraphics[width=6.5cm,angle=90]{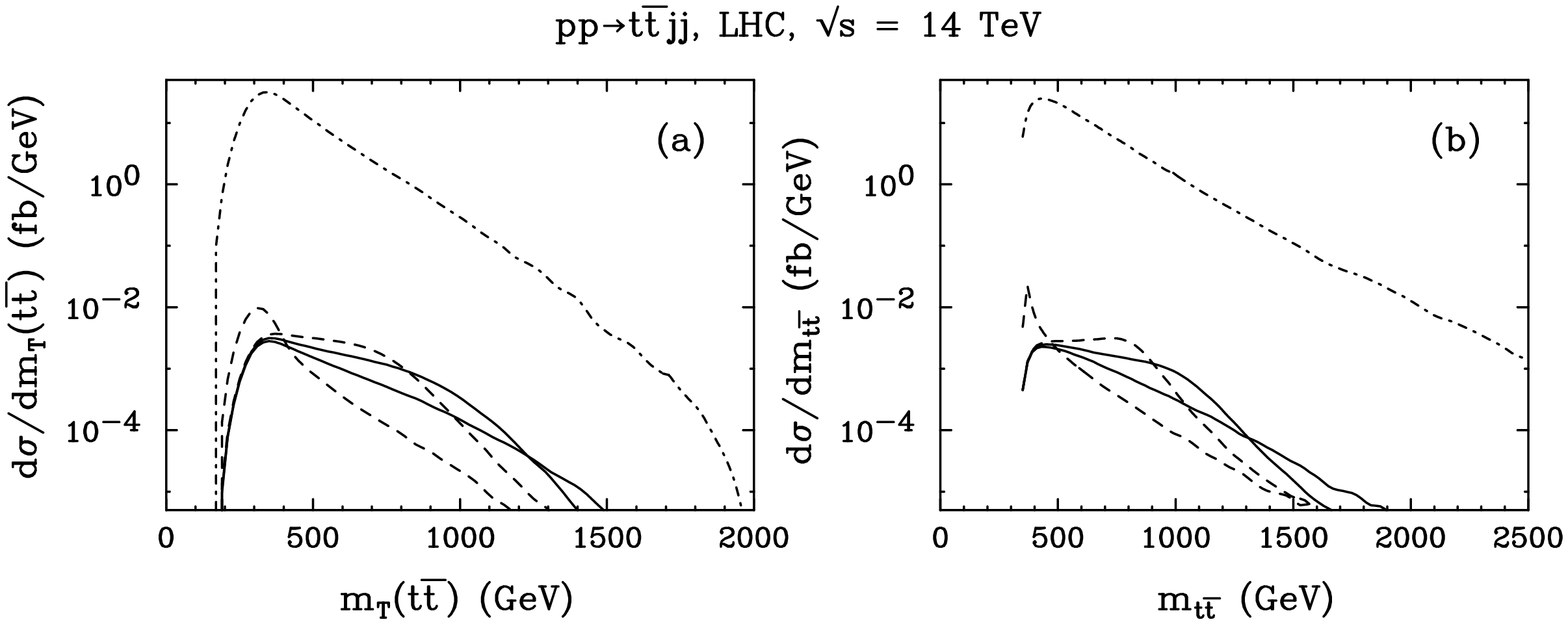}
\end{center}
\caption{Transverse mass (a) and generated invariant mass (b) of the
  top quark pair at the LHC. The SM Higgs boson signals for $M_H =
  360,800$~GeV are shown by the dashed curves, the two non-SM scalar
  cases by the solid curves, and the QCD continuum by the dot-dashed
  curve. Note that panel (b) does not represent a realistic
  observable.}
\label{fig:mtops_ls}
\end{figure}
\begin{figure}[!htb]
\begin{center}
  \includegraphics[width=6.5cm,angle=90]{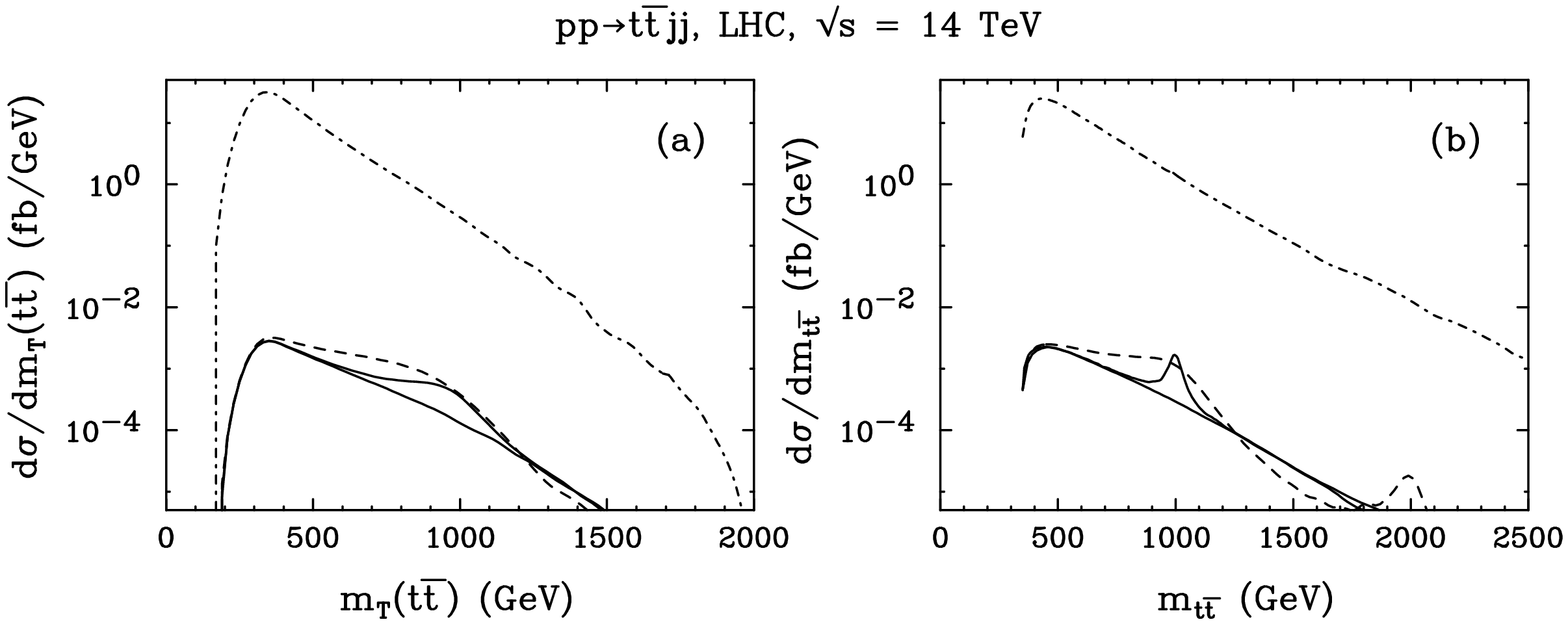}
\end{center}
\caption{Transverse mass (a) and generated invariant mass (b) of the
  top quark pair at the LHC. The non-SM heavy vectors are shown by the
  dashed curves, the non-SM scalar+vector case by the dashed curve,
  and the QCD continuum by the dot-dashed curve. Note that panel (b)
  does not represent a realistic observable.}
\label{fig:mtops_lv}
\end{figure}

There are a few obvious possibilities to pursue. The first is to cut
on the invariant mass of the forward tagging jets, a standard
technique~\cite{WBF2}. We show the various signal and QCD background
$m_{jj}^{\rm tags}$ in Fig.~\ref{fig:mtags_l}. For example, a cut of
$m_{jj}^{\rm tags} > 1$~TeV would reduce any of the signals by about
$1/4$, but the QCD background by only about $3/4$, clearly not a
significant gain. Another possibility is to look for the resonance
peak of the top quark pair due to the new scalar or vector state.
Unfortunately, because of the single leptonic top quark decay and
resulting neutrino that escapes the detector, the top quark pair
cannot be completely reconstructed. But one can determine the
transverse mass of the top quark pair, which is almost as good.

We plot $m_T(t\bar{t})$ in
Figs.~\ref{fig:mtops_ls}(a),~\ref{fig:mtops_lv}(a) for the various
signals and QCD background. It becomes immediately apparent that a
peak beyond the electroweak continuum exists only for a relatively
light state, close to $2m_t$, and the comparatively very narrow 1~TeV
vector resonance. The reasons for this are a combination of effects.
First, the very large widths of the states smear out the resonances
into the EW continuum. Second, detector effects cause mismeasurement
of the events which smear out the transverse mass peaks to the point
of indistinguishability. We verify this by plotting the generated top
quark pair invariant mass in Figs.~\ref{fig:mtops_ls}(b) and 
\ref{fig:mtops_lv}(b), which would be the ideal variable for the
signal search if it were reconstructible. There
are no visible resonance peaks for the heavy, wide states in this
distribution, although the heavy narrow states do have a peak above
the EW continuum. Third, for the 2~TeV narrow state considered in the
last model, jet separation cuts remove almost all the events, as the
highly boosted top quark decay into very narrow, massive jets. 
More detailed Monte Carlo simulations and different techniques
need to be developed in order
to discriminate these kinds of new physics events from SM backgrounds,
which is beyond the scope of the current study.

One might hope to use the upper end of a distribution, if not an
actual peak, for the more massive states. Unfortunately, the QCD
background has an extremely broad continuum distribution in
$m_T(t\bar{t})$, similar to the EW continuum, and there is almost no
event rate left in this tail, so this does not appear to be possible
at the LHC. Even for the case where some vestige of a peak exists, the
signal cannot stand out against the overwhelming QCD continuum. One is
simply at the mercy of QCD production of the same massive final state.

\begin{figure}[!htb]
\begin{center}
  \includegraphics[width=6.5cm,angle=90]{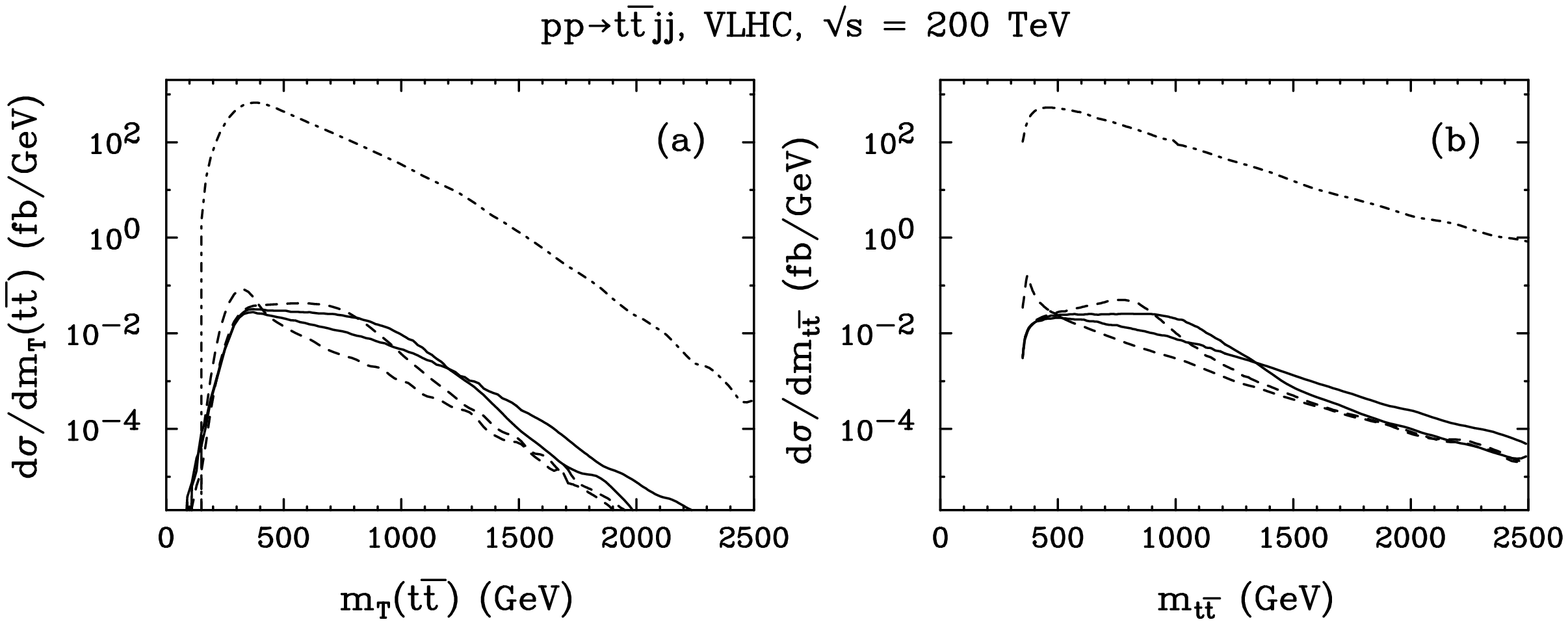}
\end{center}
\caption{Transverse mass (a) and generated invariant mass (b) of the
  top quark pair at a VLHC. The SM Higgs boson signals for $M_H =
  360,800$~GeV are shown by the dashed curves, the two non-SM scalar
  cases by the solid curves, and the QCD continuum by the dot-dashed
  curve. Note that panel (b) does not represent a realistic
  observable.}
\label{fig:mtops_vs}
\end{figure}
\begin{figure}[!htb]
\begin{center}
  \includegraphics[width=6.5cm,angle=90]{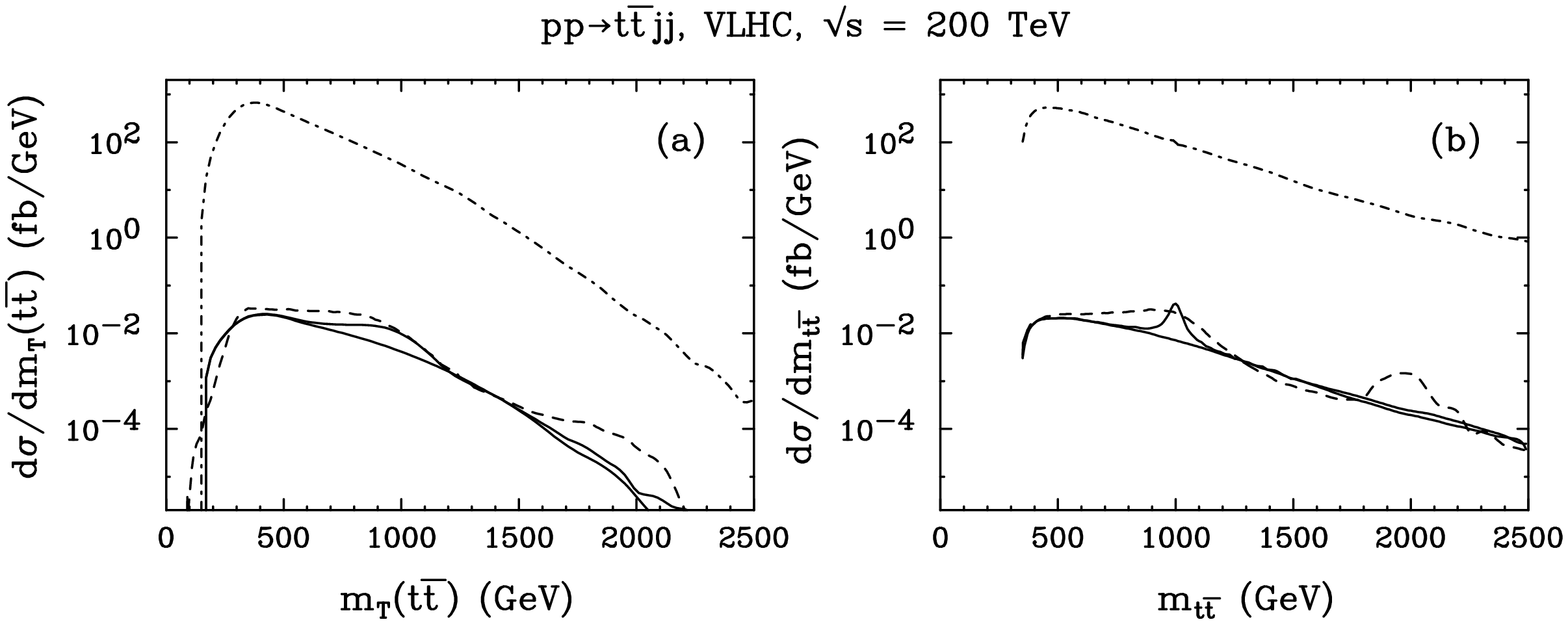}
\end{center}
\caption{Transverse mass (a) and generated invariant mass (b) of the
  top quark pair at a VLHC. The non-SM heavy vectors are shown by the
  dashed curves, the non-SM scalar+vector case by the dashed curve,
  and the QCD continuum by the dot-dashed curve. Note that panel (b)
  does not represent a realistic observable.}
\label{fig:mtops_vv}
\end{figure}

One could consider the opportunity of a VLHC instead, $pp$ collisions
at $\sqrt{s} = 200$~TeV. For these energies we increase the kinematic
cut thresholds to avoid an expected large jet multiplicity at low
$p_T$: we require $p_T(j,tag) > 40$~GeV and $p_T(b,j) > 30$~GeV. At
200~TeV the QCD continuum increases by a factor 30, to 230 pb in the
semi-leptonic decay channel. However, the SM Higgs boson signal for
example increases by only a factor 10 for $M_H = 360$~GeV, to 14.4 fb,
and a factor 14 for $M_H = 800$~GeV, to 22.0 fb. The non-SM heavy
resonances have similar cross sections, eliminating the possibility of
any gain there as well. We again show the transverse mass of the top
quark pair in Figs.~\ref{fig:mtops_vs}(a),~\ref{fig:mtops_vv}(a). As
this continuum will exhibit the same broad distribution at a much
higher energy VLHC, we conclude that the massive TOP-SEWS states and
their subsequent decay to top quarks are not likely to be a viable
channel to observe at the VLHC.

\section{Conclusion}

Given the high energy and luminosity achievable at future hadron
colliders, and the fact that the LHC will be in its mission in the
near future, it would be important to quantify to what extent the
physics of a novel electroweak symmetry breaking sector that strongly
couples to the top-quark sector can be explored.  We constructed
models for heavy scalar and vector resonances with large couplings to
top quarks and electroweak gauge bosons that respect unitarity in the
energy region probed by the LHC.  We computed the rates for the
processes $qq'\to q'' q''' t\bar t$ via gauge-boson fusion mediated by
these resonances without making approximation for the gauge bosons,
and studied whether these processes would be observable at the LHC or
the VLHC. We find that in all cases the signals are completely swamped
by QCD backgrounds. A simple way to understand this is that our models
produced in all cases event rates of the same order of magnitude of
that one obtains by simply using a heavy SM Higgs boson.
Unfortunately, the QCD background gives rates that are larger by more
than three orders of magnitude, making it impossible to extract the
signal. No sharp peak in any reconstructible distribution can be
observed as a result of smearing out from missing energy and detector
effects.

We have also considered the possible production of a vector state via
the direct $q\bar q$ annihilation. Since the direct coupling of such a
vector state to light fermions is constrained to be negligibly small
\cite{bessre}, we considered production through the $\gamma/Z-V$
mixing induced couplings as given in Ref.~\cite{bessre}.
Unfortunately, the signal cross sections turn out to be very small.
For the two cases under our consideration, we obtained the production
cross section via $q\bar q \rightarrow \gamma/Z-V \rightarrow t\bar t$
to be about 3 (130) fb for $M_V=1$ TeV, and about 0.02 (2.3) fb for
$M_V=2$ TeV at the LHC (VLHC) energies, which are too small to be
significant.

As a final note, the TOP-SEWS models under consideration often have
charged heavy scalar or vector states in addition to the heavy neutral
states. It would be desirable to observe these states via decay to a
single top quark. In this case the most obvious background is single
top quark production, which is an electroweak process. However, the
leading cross section comes from $Wb\to t$ fusion and the rate is
smaller than QCD $t\bar{t}$ production only by about a factor of three
at LHC energy \cite{scott}. Even the $t\bar{t}+jets$ production will
be a background, when some of the decay products are lost. The signal
for charged (color singlet) resonances at the TeV region does not look
more promising than that for the neutral resonant states as studied in
this paper. We therefore have not considered the charged states in
more detail.

Future hadron colliders such as the LHC or VLHC are truly
top-quark factories, with $\approx 250$ million $t\bar t$ events produced
for instance at the LHC per 300 fb$^{-1}$ luminosity.  Ironically,
this large top-quark cross section from QCD becomes the major obstacle
for studying the possible novel electroweak symmetry-breaking
physics associated with the top quark. 
In contrast, lepton colliders with a c.m.~energy of 1.5 TeV or
higher would have the potential to explore this class of electroweak
physics, and to determine the leading partial decay widths of the new 
states (which characterize the coupling strengths) statistically to 
about $10\%$ \cite{us}.

{\it Acknowledgments}: The work of T.H. was supported in part 
by the US DOE under contract No. DE-FG02-95ER40896 and in part by the
Wisconsin Alumni Research Foundation. The work of G.V. was supported
in part by DOE under contact number DE-FG02-01ER41155. We would also 
like to thanks the organizers of the Snowmass 2001 workshop, where 
part of this work was conducted.


\end{document}